\documentclass[12pt,titlepage]{article}

\usepackage{graphicx}
\usepackage{endfloat}
\usepackage{amsfonts}
\usepackage{mychicago}
\usepackage{subfigure}
\usepackage{genetics_manu_style}
\usepackage{amsmath, amssymb, graphicx, epsfig,subfigure,wrapfig}

\newcommand{\bx}{\mathbf{x}}
\newcommand{\bw}{\mathbf{w}}

\newcommand{\cL}{{\cal L}}

\title{Estimating heterozygosity from a low-coverage genome sequence, leveraging data from other individuals sequenced at the same sites  \\
\protect\small An Investigation Submitted to {\em Genetics}}

\author{Katarzyna Bryc\thanks{Department of Genetics, Harvard Medical School,
Boston, MA 02115} Nick Patterson\thanks{Broad Institute of Harvard and MIT, 7 Cambridge Center, Cambridge, MA 02142} David Reich$^*$}

\renewcommand{\CorrespondingAddress}{Department of Genetics \\
    Harvard Medical School \\ 
    77 Ave. Louis Pasteur\\ 
    Boston, MA 02115 \\ 
    (617) 432-1101 (ph.)  \\
    (617) 432-6306 (fax)  \\ 
    \texttt{kbryc@genetics.med.harvard.edu} \vfill}
\renewcommand{\RunningHead}{Estimating heterozygosity}
\renewcommand{\CorrespondingAuthor}{Katarzyna Bryc}
\renewcommand{\KeyWords}{Heterozygosity, Low-coverage sequence data}

\begin{document}

\maketitle

\begin{abstract}
High-throughput shotgun sequence data makes it possible in principle to accurately estimate population genetic parameters without  confounding by SNP ascertainment bias. One such statistic of interest is the proportion of heterozygous sites within an individual's genome, which is informative about inbreeding and effective population size. However, in many cases, the available sequence data of an individual is limited to low coverage, preventing the confident calling of genotypes necessary to directly count the proportion of heterozygous sites. Here, we present a method for estimating an individual's genome-wide rate of heterozygosity from low-coverage sequence data, without an intermediate step calling genotypes. Our method jointly learns the shared allele distribution between the individual and a panel of other individuals, together with the sequencing error distributions and the reference bias. We show our method works well, first by its performance on simulated sequence data, and secondly on real sequence data where we obtain estimates using low coverage data consistent with those from higher coverage. We apply our method to obtain estimates of the rate of heterozygosity for 11 humans from diverse world-wide populations, and through this analysis reveal the complex dependency of local sequencing coverage on the true underlying heterozygosity, which complicates the estimation of heterozygosity from sequence data. We show filters can correct for the confounding by sequencing depth. We find in practice that ratios of heterozygosity are more interpretable than absolute estimates, and show that we obtain excellent conformity of ratios of heterozygosity with previous estimates from higher coverage data.
\end{abstract}

\section{Introduction}

Heterozygosity, or the fraction of nucleotides within an individual that differ between the chromosomes they inherit from their parents, is a crucial number for understanding human variation. Estimating this simple statistic from any type of sequence data is confounded by sequencing errors, mapping errors, and imperfect power for detecting polymorphisms. Obtaining an unbiased estimate is especially difficult for ancient genomes where the sequences have a higher error rate, or in cases of low-coverage sequence data where there is low power to detect heterozygous sites, or for hybrid capture where there may be additional biases due to the oligonucleotides used for fishing out sequences of interest.

Several methods for estimating individual heterozygosity have been proposed \cite{JohnsonSlatkin2006,Hellmann2008,Jiang2009,Lynch2008,Haubold2010}. For an overview of these methods see \cite{Haubold2010}. Haubold \textit{et al.} 2010 \cite{Haubold2010} describe \textit{mlRho}, an implementation of a method that jointly infers $\theta$, the scaled mutation rate, and $\rho$, the scaled recombination rate for a shotgun sequenced genome. However, they examined performance of their method at 10X coverage and a small sequence error rate of $4 \times 10^{-4}$, which is about four times lower than encountered currently in real data \cite{Shendure2008}. We developed a method that estimates the heterozygosity for an individual of interest by leveraging the genome-wide joint information across sequence reads from a panel of individuals.  The advantage of leveraging the panel of individuals in our method is that it enables learning of the empirical distribution of alleles at heterozygous and homozygous positions, a distribution that encapsulates sequencing errors and the non-Bernoulli sampling of each allele at a heterozygous SNP. This allows one to disentangle the rate of heterozygosity from sequencing errors and other biases, and does not require explicit modeling of these platform--, batch--, and genome--specific (frequently unknown and unestimatable) error processes. As a result of including the allele or genotype information at other individuals, our method gains robustness to any unknown error sources that may also be present within the data. 

We use an Expectation-Maximization (EM) algorithm to estimate the most likely distribution of counts across the unknown underlying genotypic states, from which we obtain an estimate of the proportion of loci that are heterozygous in the target individual. An advantage of this method is that it returns an unbiased and accurate estimate of heterozygosity even when the individual has low sequence coverage. Our method learns the distribution of alleles directly from the sequence read data, and does not require modeling demographic relationships among the individuals nor genotype calls from the sequence reads. We validate our EM method on 1 Gb of simulated sequence data of 5X, 10X, and 20X coverage, and find that our method performs well at estimating the true heterozygosity even when the sequence error rate is extreme and mean coverage is low. As an empirical validation of the ability of our method to perform well on low-coverage datasets, we test our method on real high-coverage (~30X) sequencing data, which we subsample to lower coverage, and verify that our estimates are consistent. In particular, we show that applying our method to a lower-coverage subsampling provides the same estimates of heterozygosity as those obtained on higher-coverage data, and are concordant with estimates of heterozygosity from other methods.

We apply our method to obtain estimates of heterozygosity for 11 individuals from many world-wide human populations, from \cite{Meyer2012}. Our finding underscores the need to compare ratios of heterozygosity across fixed genomic regions to infer the relative rates of diversity among individuals.

\section{Materials and Methods}
We apply our method to read data at sites with a target minimum coverage (for example, $\geq$ 5X coverage) for the  sequenced diploid individual of interest, aligned to some reference genome of known sequence. We also use sequence read data from $n$ other individuals likewise aligned to the reference.

Let $a$ be the unknown diploid genotype of our target individual at some position in the genome, and $c$ be the aligned reference allele. Then the allele distribution in other individuals will depend on $g = (a, c)$. Let ${\bf x} =(x_1, x_2, \ldots, x_n)$ be the vector of alleles by taking one randomly sampled read from each of the $n$ individuals.  Let $\bw$ be the observed alleles from the reads for our individual. Both ${\bw}$ and ${\bf x}$ are observed quantities for a given position in the genome for our individual, and we are interested in modeling the joint probability of  ${\bw}$, ${\bf x}$ as the sum of the joint probabilities conditional on $g$. 

We assume conditional independence of $\bw, \bx$ on the true unobserved genotype $g$. This assumption holds if the allele frequency spectrum of the panel of individuals depends only on the true underlying genotypic state of our individual, and not the allele counts we observe, and likewise the allele count distribution depends only on the true underlying genotypic state and not on the the alleles observed in the other individuals. From this conditional independence property, we then derive:
\begin{eqnarray}
P(\bw, \bx | g) &=& P(\bw |g)P(\bx | g) \\
P(\bw, \bx) &=&  \sum_{g}P(g)P(\bw, \bx|g) \\
&=& \sum_{g}P(\bw | g)P(\bx |g)P(g)\\
P(\bw, \bx, g) & = &P(\bw | g)P(\bx | g) P(g)
\end{eqnarray}
which will later provide the leverage to infer the most likely values for the above probabilities,  including $P(g)$, which gives us the genomic rate of heterozygosity.

For every site which has both sufficient coverage in our individual and for which we have complete information of the panel, we add this site to the corresponding bin of observed alleles $\bw$  and  panel $\bx$. This full matrix would be inconveniently large, so to simplify the data matrix of counts, we polarize our allele counts with respect to the reference, restricting to bi-allelic SNPs, which constitute the majority of sites. We denote the reference allele as $0$, and allow only a single other variant per site, summarizing the observed alleles from the reads by the number of non-reference alleles. Thus, we denote genotypes as $g \in \{0,1,2\}$, which we will refer to as the homozygous ancestral, heterozygous, and homozygous derived states, respectively. If, for example, we consider only sites with a coverage of 4, then $\bw \in \{(4,0), (3,1), (2,2), (1,3), (0,4)\}$. We can also easily represent $\bx$ as a vector of 0's and 1's referring to reference or non-reference allele present in the randomly sampled reads, for example, $\bx = (0,1,1,0,1)$ where the length of $\bx$ is determined by the number of individuals we sample. 

We create a count matrix $N$ of dimension $||\bw || \times ||\bx ||$ corresponding to the number of observed sites with each particular combination of $\bw$ and $\bx$.  The counts of the number of loci where the individual is $\bw$ and the panel of individuals individuals comprise the alleles $\bx$ is represented by the corresponding row and column entry in the matrix $N$.

From the matrix $N$ we estimate the true values of $P(g), P(\bw | g)$, and $P(\bx | g)$ using the EM algorithm.  Let $Y_{obs}$ be the observed counts of alleles in the matrix $N_{\bw, \bx}$. Let $Y_{mis}$ be $N_{\bw, \bx , g}$, the missing or unobserved counts of the alleles with the true parameter state $g$.  Then the likelihood of the data is:
\begin{equation} 
\cL = 
\sum_{\bw , \bx} N_{\bw , \bx} \log P(\bw ,\bx)
\end{equation} 
If the hidden variable, $g$, corresponding to the true underlying genotypic state were observed, the log-likelihood would be:
\begin{equation} 
\cL' = 
\sum_{\bw , \bx, g} N_{\bw , \bx, g} \log P(\bw ,\bx, g)
\end{equation} 
But this would require fitting $|| g ||$ different parameters per observed data point (i.e., count entry of $N_{\bw, \bx}$). This would require fitting three times as many parameters as there are data points. However, by relying on our conditional independence from equation (4) above we can reduce the number of parameters to be fitted from the data.

By EM theory, the $Q$ function $Q(P, \hat P)$ is given by:
\begin{eqnarray*} 
Q(P, \hat P) & = & 
E_{post} L' (\hat P)  \\
& = & 
\sum_{\bw , \bx, g} \hat N_{\bw , \bx, g} \log \hat P(\bw ,\bx, g)
\end{eqnarray*} 
where $ \hat N_{\bw , \bx, g} $ is the expected value of $ N_{\bw , \bx, g} $, 
which in our case derives from the multinomial distribution, under the posterior distribution calculated  with the old parameters $P$.

The estimates  for $\hat P$ that maximize $Q$, also derived from the MLE estimates for the multinomial distribution, are: 
\begin{eqnarray*}
\hat{P}(\bw | g) & =  & \frac{\sum_\bx \hat N_{\bw ,\bx ,g}}{\sum_{\bw ,\bx } \hat N_{\bw ,\bx ,g}} \\
\hat{P}( \bx |g) & = & \frac{\sum_\bw \hat N_{\bw ,\bx ,g}}{\sum_{\bw ,\bx } \hat N_{\bw ,\bx ,g}} \\
\hat{P}(g)& = & \frac{\sum_{\bw , \bx } \hat N_{\bw ,\bx ,g}}{N} 
\end{eqnarray*}

Further, by Bayes theorem this expands to,
\begin{eqnarray*}
\hat{N}_{\bw ,\bx ,g } & = & N_{\bw ,\bx} \cdot \frac{\hat{P}(\bw ,\bx ,g)}{\hat{P}(\bw ,\bx)} =  N_{\bw ,\bx} \cdot \frac{\hat{P}(\bw  | g)\hat{P}(\bx  | z)\hat{P}(g)}{\sum_z\hat{P}(\bw  | g)\hat{P}(\bx  | g)\hat{P}(g)}
\end{eqnarray*}

By basic EM theory these re-estimated values of $\hat P$ will generate a non-decreasing sequence of values for the log likelihood $\cL$. Finally, we obtain the parameter  of interest $\hat{P}(g = 1)$ after convergence.

\subsection{Implementation}

In practice, without constraining the parameters $\hat{P}(\bw | g)$ we reach local but not consistently global likelihood maxima, which do not necessarily correspond to the genotypic state parameters we wish to obtain. To improve the ability of the EM to achieve global maxima, we fit Beta-Binomial distributions (effectively an over-dispersed Binomial distribution) to the probabilities of number of non-reference alleles $P(\bw | g)$ for each possible genotypic state $g$. This constraint, as well as the choice of reasonable starting parameters for the EM initialization, in practice improves our convergence to global maximum corresponding to the homozygous ancestral, heterozygous and homozygous derived genotypic states.

Like the Beta distribution, the MLE estimates for the Beta-Binomial distribution do not have a closed form, though they can be found using direct numerical optimization [such as a fixed-point iteration or a Newton-Raphson iteration]. However, instead, we estimate the two parameters ($\alpha, \beta$) using MOM estimators for the Beta-Binomial, by setting:
\begin{align*}
\hat{\alpha} &=  \frac{(n - \bar{x} - s^2/ \bar{x})\bar{x}}{(s^2/\bar{x} + \bar{x}/n - 1)n}\\
\hat{\beta} & = \frac{(n - \bar{x} - s^2/\bar{x})(n-\bar{x})}{(s^2/\bar{x} + \bar{x}/n - 1) n}
\end{align*}


In the case of under-dispersed data, it is possible to obtain MOM estimates that are invalid. Though unlikely to occur in the read data, for this contingency, we instead fit a Binomial distribution to the data.

There are several challenges in implementing the EM for our problem. The first is that the likelihood is poorly defined when any of the parameters we are interested in estimating approach 0, as then the likelihood also goes to zero. So to avoid this situation, we add a ``prior'' $\epsilon$ to the likelihood calculation which adds a small count value in the step calculating the parameters to avoid probabilities reaching 0. So instead, we calculate the posterior:
\[ L' = \hat{N}_{p,x,z} log(P(p,x,z)) + L' \]
 rather than the maximum likelihood estimation, hence, we obtain a MAP (maximum a posteriori) estimate, which is a Bayesian method that incorporates a prior over the distribution to be estimated (in this case, a small uniform prior). We choose a small prior (less than in total counting one site across all possible matrices) that does not impact our estimates. In general, our estimates are robust to choice of this prior, within a range examined of 1e-10 to 1e-50, and we will continue to refer to our method as an EM implementation though in fact we use a non MLE method. In effect, our equations for each step remain the same, except that in the M-step of the EM (where we estimate the parameters) we instead estimate the MAP using the prior. Specifically, we estimate:
\[Posterior = \sum_{i,j} ( N_{i,j} \cdot log(\sum_{z} P_{i,j,z}) + \epsilon \cdot \sum_{z} log(P_{i,j,z}) ) \]
In practice, we set $\epsilon$ to 1e--20 which does not alter estimates of the probabilities while preventing ill-defined likelihoods.

As with all likelihood calculations, our probabilities approach very small numbers. To avoid numerical error due to underflow of small likelihoods and parameter estimates, we implement the EM storing all probabilities and likelihoods in the log form.

Lastly, likelihood maximization occurs on an arbitrary base, so to avoid numerical issues due to any remaining underflow of the likelihood calculation, we compute a factor $F$ at the start of the EM. For each iteration, we compute the likelihood of the data minus this constant factor, which is a standard practice and does not affect the computation of the maximum. This is equivalent to calculating the log odds:
\begin{eqnarray*}
L &= & \left( \sum_{i,j} N_{i,j}\cdot log(P_{i,j}) \right) - \left(F\right)\\
&=& \left( \sum_{i,j} N_{i,j}\cdot log(P_{i,j})\right) - \left( \sum_{i,j} N_{i,j}\cdot log(F_{i,j})\right)\\
&=&  \sum_{i,j} N_{i,j}\cdot (log(P_{i,j}) - log(F_{i,j})) \\
&=& \sum_{i,j} N_{i,j}\cdot log(\frac{P_{i,j}}{F_{i,j}})
\end{eqnarray*}

for some constants $F_{i,j}$. In practice, we set $F_{i,j}$ to be the likelihood at initialization of the EM. We then iterate the EM until both the change in parameters and the change in the likelihood is smaller than our chosen threshold, which in practice we set as 1e--50.

It should be noted that any form for tallying read counts may be used, including the allele profile used in \cite{Haubold2010}; our choice of was motivated by a choice of dimensionality that is a compromise between simplicity and capturing relevant information. Our method is highly generalizable to any choice of count data, and could be implemented assuming that a reasonable starting position for the EM could be proposed, such that the iterations are likely to converge to a global maximum corresponding to the genotypic states.

\subsection{Proof of principle 1: Application to simulated data}
We generated simulated sequence data and applied our EM method for estimating heterozygosity to assess the accuracy of our estimation procedure.

\subsubsection{Generating coalescent simulations}
We generated 100 replicate datasets of sequence data using MaCS \cite{Chen2009}. Each replicate dataset contains 10 chromosomes of length 100Mb for a total of 1Gb of sequence, for 1 chimpanzee chromosome, 7 African chromosomes, 5 European chromosomes, and 5 East Asian chromosomes,  using demographic parameters fitted by \cite{Gutenkunst2009}. We include a chimpanzee chromosome assuming a constant population size of 50,000 individuals and a split time from humans of 6Mya, using the same generation time as humans of 25 years per generation. 


\subsubsection{Adding simulated error}  We simulate sequence data from the true genotypes by adding errors to reads. First, for all variable loci in the target individual, we randomly choose which allele is on a read, then adds errors to the each read (with high error rate of 0.002) to generate the total number of derived reads for the individual at the locus out of the total sequencing depth. For each other sequenced chromosome, we add errors with a lower error rate of 0.0001 (since we assume the panel is composed of higher-quality genomes), then add a count for the final simulated locus in the appropriate hash bin. Lastly, for each invariant locus, we add errors to the target individuals reads, and at a lower rate, adds errors to the other sequenced chromosomes, and input these counts into the hash bin. With an error rate of 0.001 we add errors to the chimpanzee chromosome which inverts the ancestral and derived reads.

\subsection{Proof of principle 2: Downsampling high-coverage genomes}

To assess the efficacy of our method at lower coverages on real sequence data, we obtain estimates of heterozygosity for a San individual sequenced to higher coverage using  Illumina's Genome Analyzer IIx next-generation sequencing technology, which we then downsample to varying levels low coverage. We use this dataset of sequence reads to explore the ability of our method to perform on low-coverage sequence data, and the lower bound of coverage at which we are able to obtain accurate estimates of heterozygosity. We compare the performance of our method to the estimate of $\theta$ obtained from \textit{MlRho} \cite{Haubold2010}.

\subsection{Application to 11 world-wide human genomes}

We align sequence data from 11 human genomes from world-wide populations and an archaic Denisovan genome to the chimpanzee reference genome to avoid introducing human-reference-population biases. For details on the populations, samples and the sequencing performed, see \cite{Meyer2012}. We generate a counts matrix for each of the genomes  using a panel generated by a single read sampled from each of the other 11 genomes.

We include only sites where there is a chimpanzee reference allele, exclude sites where two or more non-reference bases are equally present or if there are more than 5 reads showing a third (non-variant and non-reference) allele. We also exclude CpG sites, as well as sites where any individual from the panel has no coverage or sites that have insufficient coverage for the target individual. 

To demonstrate the relationship between sequencing coverage and the true rate of heterozygosity of different regions, we generate count data for each bin of 5X coverage ranging between 5X and 50X, where for each bin dataset, we only include sites where the coverage of the individual falls within the target range. We downsample coverage at each bin (where possible) to 5X, 10X, and 20X, and compare results stratified by downsampling, as well as by genomic coverage.

Lastly, we produce estimates for the 11 present-day genomes and the archaic Denisova genome on a fixed set of sites, and compare to previous estimates for these samples \cite{Meyer2012}.

\section{Results}

\subsection*{Simulation results}

\begin{figure}[htbp]
\begin{center}
\includegraphics[width=1\textwidth]{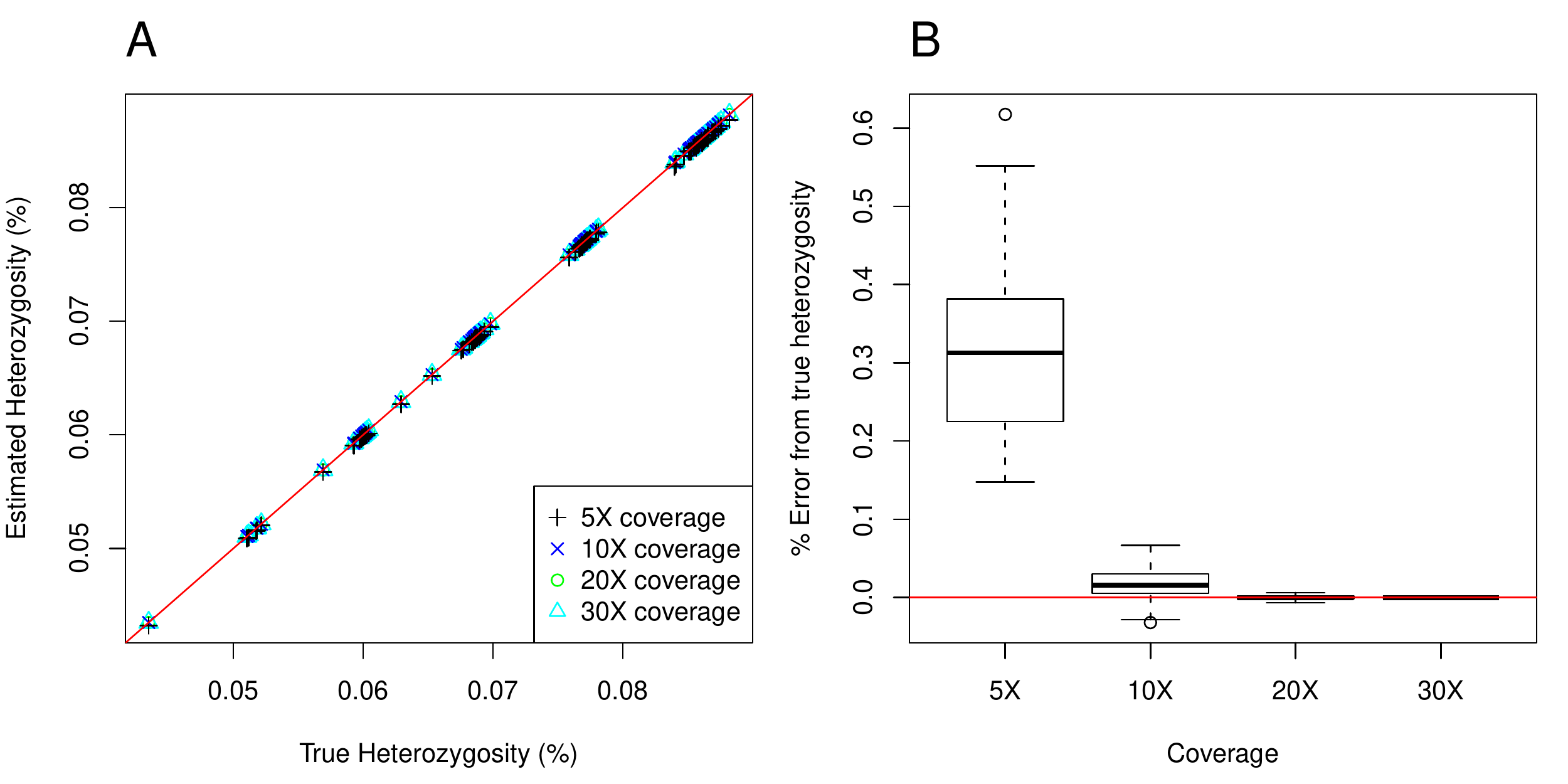}
\caption{True versus estimated rates of heterozygosity for 100 simulated read datasets. \textit{A:} Each dataset has been downsampled to different coverage levels, denoted by symbol color and shape. The red line corresponds to True=Estimated, or perfect estimation of heterozygosity. \textit{B:}   Run-by-run differences between true and estimated heterozygosity rates, stratified by downsampling coverage. The y-axis shows the percent error from the true value of heterozygosity. }
\label{simulationResults}
\end{center}
\end{figure}

We  obtain accurate estimates of heterozygosity across a variety of coverage levels (5X to 30X) (see Figure \ref{simulationResults}). We note a tiny bias of 0.3\% (in relative terms) from the true rate for 5X coverage read data, but with higher coverage this bias goes to zero.

\subsection*{Downsampling results}

\begin{figure}[htbp]
\begin{center}
\includegraphics[width=0.8\textwidth]{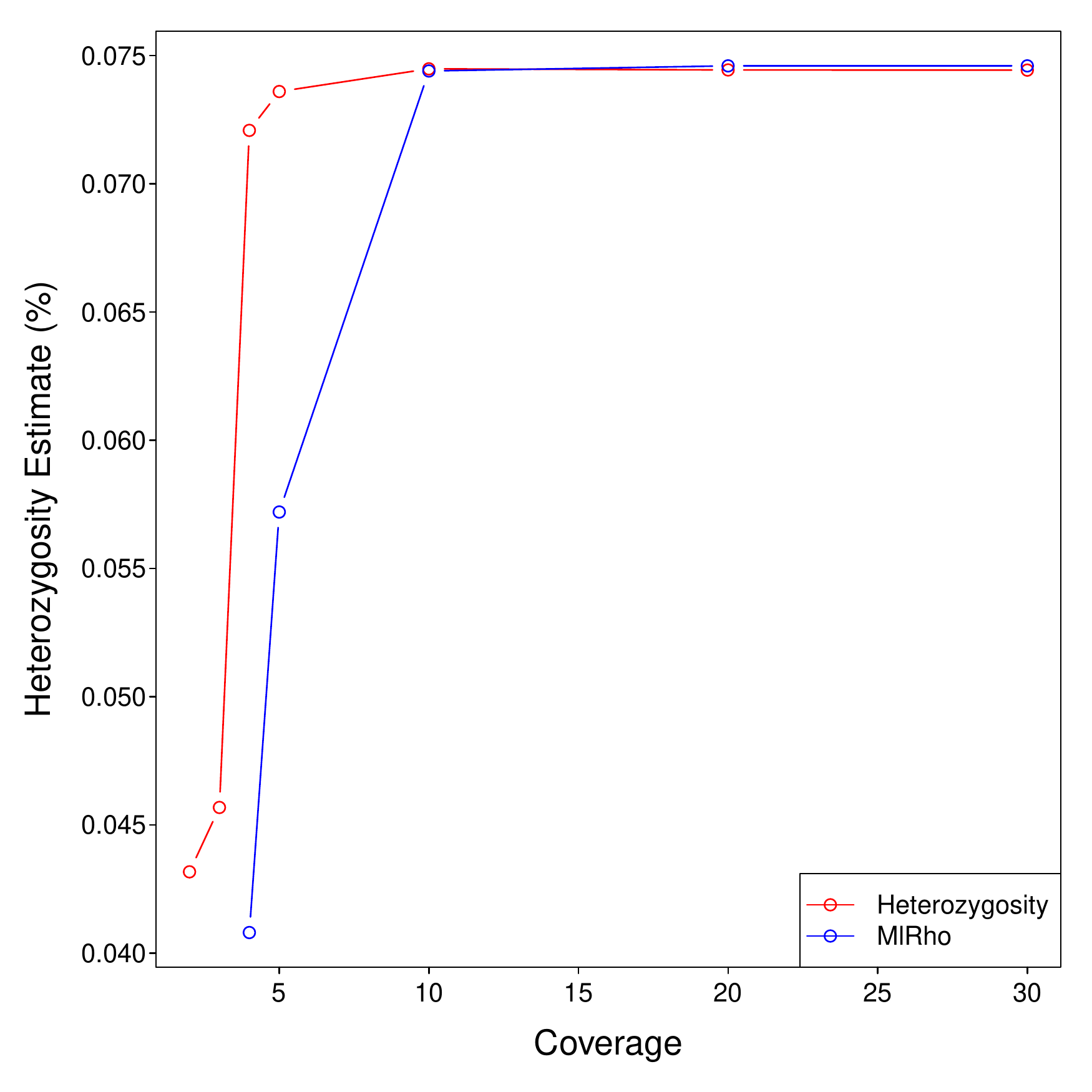}
\caption{Our EM Heterozygosity estimates (red) and \textit{MlRho} estimates on the regions of a San individual genome sequenced to 30-45X, and experimentally downsampled. At higher coverage, both methods converge to an estimate of $7.45\times10^{-4}$. We note that our estimates for 4X and 5X coverage are much more accurate than \textit{MlRho}. Results for less than 4X coverage were not possible to obtain from \textit{MlRho}.   }
\label{downsamplingResults}
\end{center}
\end{figure}

Figure \ref{downsamplingResults} illustrates that our EM estimation method and \textit{MlRho} give consistent estimates of heterozygosity for the HGDP San individual starting at about 10X coverage and higher. However, at lower coverage (about 4X) our method significantly outperforms \textit{MlRho}, giving a slightly biased, but nearly convergent, estimate.


\subsection*{Heterozygosity estimates for 11 present-day and Denisovan  genomes}

\begin{figure}[htbp]
\begin{center}
\includegraphics[width=1\textwidth]{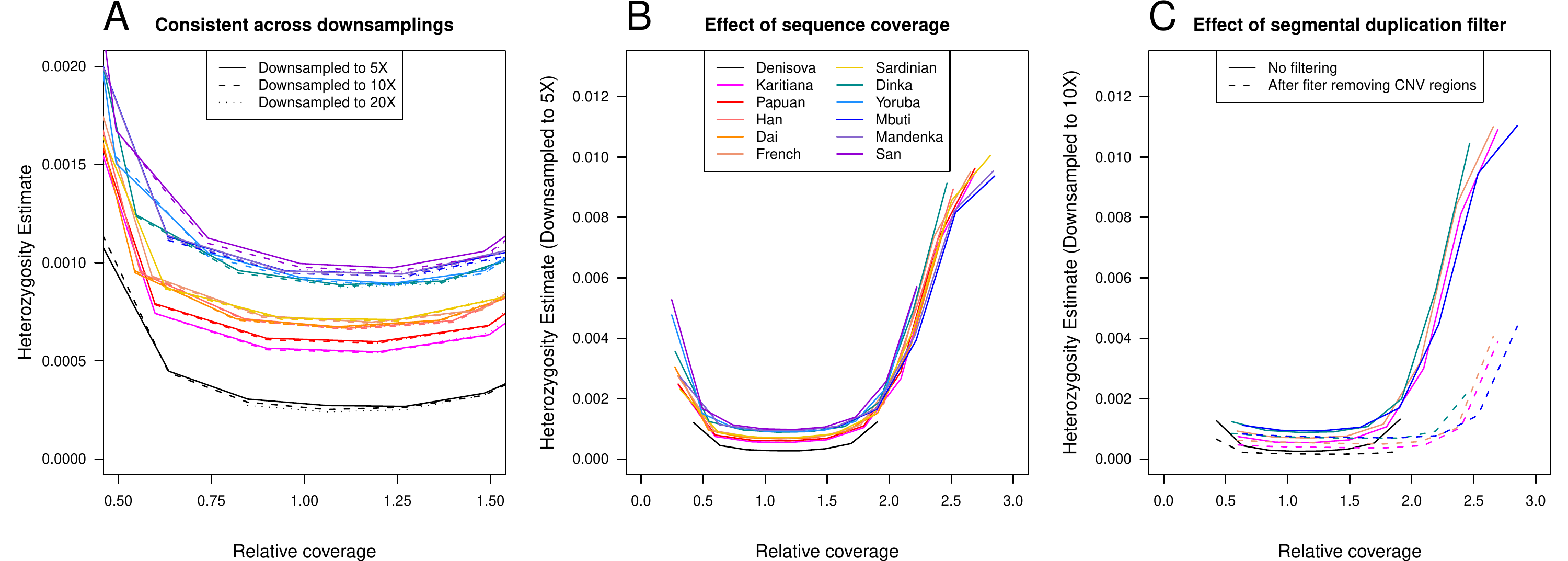}
\caption{Estimates of heterozygosity for each of the 11 present-day human genomes and Denisova, where each indvixual is denoted by a unique color. Relative coverage is defined as the lower bound of the sequencing bin, divided by the mean sequencing depth for the individual. \textit{A:} Heterozygosity estimates are consistent across downsampling levels. Downsampling to 5X, 10X, and 20X levels is denoted by line type. Each individuals is denoted by line color. \textit{B:} All individuals show an increase in estimated heterozygosity at higher  (and lower) relative coverage. \textit{C:} Effect of removing known regions with segmental duplications. Estimates of heterozygosity are shown for a sample of five of the individuals. Without filtering, estimates for each bin are shown with solid lines. After exclusion of regions within known CNV and segmental duplications, the heterozygosity estimates display a flatter distribution (dotted lines).}
\label{hetByDownAndCov}
\end{center}
\end{figure}

We present our initial estimates of heterozygosity, downsampled to three different depths, for each sequencing coverage bin (normalized by individual mean sequence coverage) in Figure  \ref{hetByDownAndCov}A. Our estimates of heterozygosity are consistent, independent of count matrix (i.e., downsampling) size, as would be expected from our simulated downsampling results shown in Figure \ref{downsamplingResults}. However, we find a strong signal that the estimates of heterozygosity are correlated to sequencing coverage of the region. We note that this is not an artifact of the larger amount of data available at higher coverage, since each bin is calculated after being downsampled to the same depth. Instead, the U-shaped curves in Figure~\ref{hetByDownAndCov}B indicate that the apparent next-generation sequencing coverage is dependent on properties of the underlying genomic sequence. In particular, we find that regions of lower coverage and higher coverage (relative to the mean sequencing depth) show higher heterozygosity.

We witnessed increased heterozygosity at regions of higher coverage, which we suspected was due to perceived genetic diversity due to cryptic segmental duplications. To explore this hypothesis, we restricted our analyses to regions of the genome which have been identified as unlikely to contain segmental duplications, available on the Eichler Laboratory website (\texttt{http://eichlerlab.gs.washington.edu/database.html}). We find that this filter strongly reduces the effect of regions with likely segmental duplications on our estimates of heterozygosity (Figure \ref{hetByDownAndCov}C), confirming that unidentified segmental duplications, which result in a net higher sequencing coverage of the region, result in an high estimate of heterozygosity for such regions. Roving these regions with known segmental duplications reduces this effect at regions with higher sequencing coverage. However, the increase in heterozygosity at higher coverage still is present even after this correction (see Figure \ref{hetByDownAndCov}C), suggesting that there may be further individual or population-specific duplications remaining.

\begin{figure}[htbp]
\begin{center}
\includegraphics[width=0.8\textwidth]{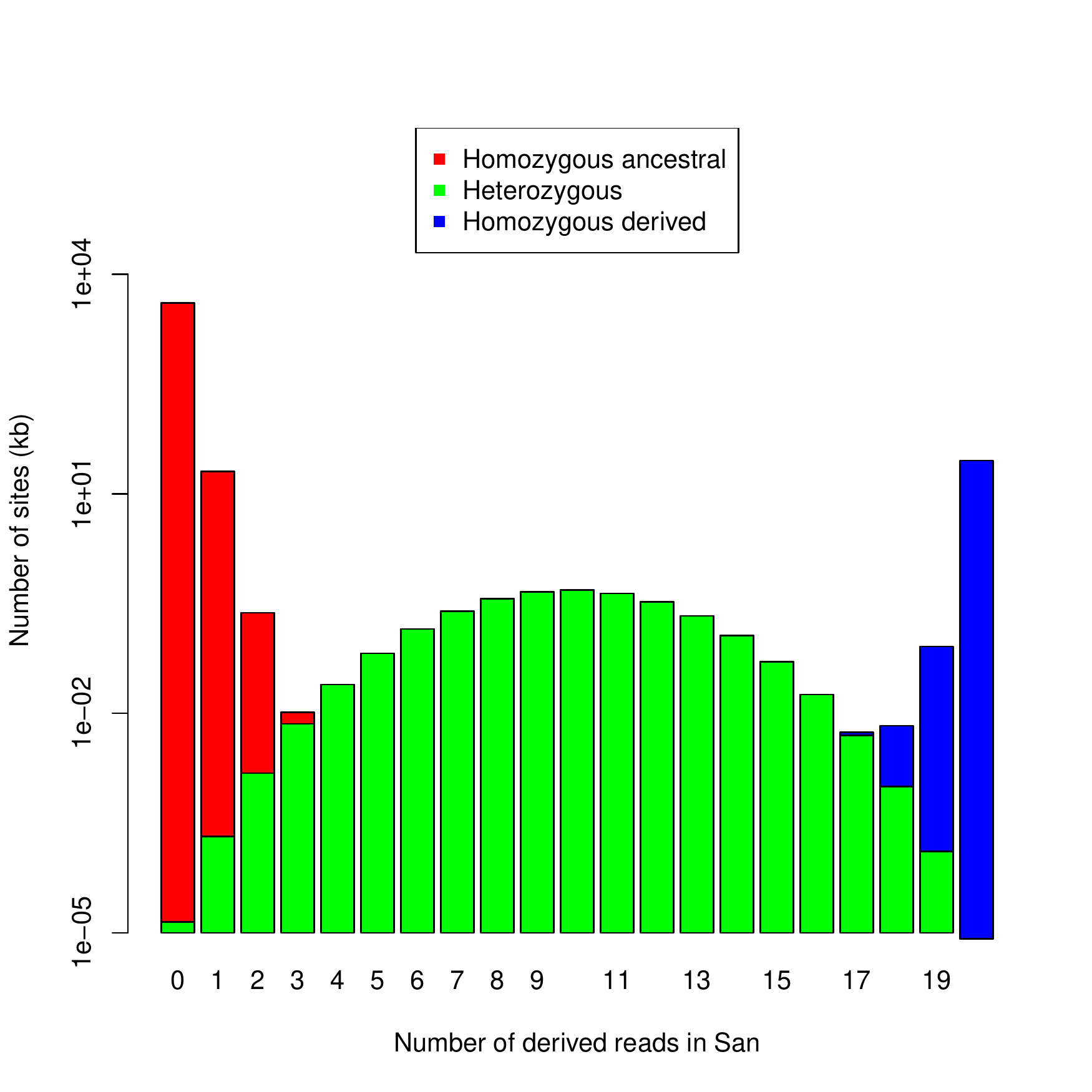}
\caption{Inferred distribution of homozygous ancestral (red), heterozygous (green), and homozygous derived (blue) sites for the San HGDP individual. The y-axis is presented on a log scale.}
\label{hetDistribution}
\end{center}
\end{figure}

Using only data that passed the segmental duplication filter, we obtain estimates for the sequenced genomes on the same set of regions, restricting to regions with sequencing coverage between 20X and 40X. Using the EM, we estimate the total genome-wide fraction of heterozygosity for each individual, and we also can extract estimates of the allelic distribution of heterozygous and homozygous sites (Figure \ref{hetDistribution}). We present the absolute estimates we obtain in Table \ref{relativeHet}, as well as the ratio of heterozygosity in the Denisova genome relative to the other individuals.  We find the highest estimates of heterozygosity for the San African individual, and next highest estimates of heterozygosity for other African individuals from Mandenka, Yoruba, Mbuti, and Dinka populations.  The next highest levels of heterozygosity are in individuals from European populations (French, Sardinian), followed by East Asian populations (Dai, Han). We find the lowest estimates of heterozygosity in the individuals from Melanesia (Papuan) and from a Native American population (Karitiana). 
\begin{table}[htdp]
\begin{center}
\begin{tabular}{lcc} \hline
Individual & Heterozygosity estimate (\%)  & Ratio \\ \hline \hline
Denisova &	0.0165 & -- \\ \hline
San &  0.0721 & 23\%\\
Mandenka & 0.0686 & 24\% \\
Yoruba & 0.0649 & 25\%\\
Mbuti & 0.0657 & 25\% \\
Dinka & 0.0635 & 26\% \\
Sardinian & 0.0490 & 34\% \\
French & 0.0473 & 35\% \\
Dai & 0.0465 & 35\% \\
Han & 0.0454 & 36\% \\
Papuan & 0.0386 & 43\% \\
Karitiana & 0.0353 & 47\% \\ \hline
\end{tabular}
\end{center}
\caption{Heterozygosity estimates for the 11 present-day individuals from world-wide populations and Denisova. The ratio presented is the relative heterozygosity in the Denisova genome as a percentage of that found in the present-day individual.  }
\label{relativeHet}
\end{table}

\section{Discussion}

We have shown that our heterozygosity estimation method both performs well in low coverage simulated sequence data and provides consistent estimates on real low-coverage data downsampled from higher coverage. In particular, our method outperforms other methods on data that has been sequenced at less than 10X coverage, and provides reasonable estimates for as low as 4X coverage.

Our estimates for 11 world-wide human genomes and the archaic Denisovan genome provide important insights into the distribution of heterozygosity across human populations. Furthermore, our results show that estimates of heterozygosity are strongly affected by genomic properties such as copy-number variability, and these properties affect sequencing coverage. Hence, we show that the heterozygosity is not independent of sequencing coverage even within one genome, and is elevated in both regions with low coverage (relative to the mean sequencing depth) as well as regions with high coverage.  This is an unexpected result if one assumes a the ``Lander-Waterman'' Poisson distribution of read depth \cite{LanderWaterman1988,Weber1997}. Furthermore, even after excluding regions with known copy-number variable regions, an increase in heterozygosity is still present at the more extreme levels of sequence coverage, suggesting that other correlations of sequence diversity with coverage, or possibly  individual-specific segmental duplications, still remain.  Implications from our result suggest that using the higher tail of sequencing coverage for population genetic inference may result in a biased set of genomic regions with selectively higher heterozygosity, possibly due to population and individual segmental duplications.

Our absolute estimates of heterozygosity are lower than those reported for these genomes in other papers using other methods \cite{Meyer2012}. The absolute estimates are notably lower, but consistent with the \textit{relative} diversity previously documented in these populations, and with other patterns of genetic diversity such as decay of linkage disequilibirium \cite{Jakobsson2008}.
Because the regions of the genome that pass our filters (and in particular, the copy-number variable filter) are likely to be lower in complexity and substantially biased towards lower diversity due to alignment biases, the lower absolute values of heterozygosity are expected.  However, our relative heterozygosity estimates are consistent with previously documented levels of genetic diversity, with African populations showing highest levels of diversity, and decreasing levels with distance away from Africa \cite{Li2008,Jakobsson2008}. Our estimates confirm previous findings that the archaic Denisovan genome shows substantially lower levels of heterozygosity than any of the other present-day populations, with only a fraction of the rate of heterozygosity.

More generally, we emphasize that absolute heterozygosity is not a well-defined quantity in the analysis of genomic data, as it strongly depends on the particular filters that are used to select the regions being analyzed, and may be an implausible concept in highly repetitive regions (such as centromeres and telomeres) and copy-number-variable regions.  The absolute value of heterozygosity can vary based on the regions chosen to be examined, but the relative heterozygosity estimates or ratios among individuals (using the same regions and filters) are consistent. Hence, in practice, heterozygosity estimates are most meaningful when viewed as relative ratios among individuals for the same regions of the genome, and not as absolute values inherent to diploid genomes. 

\section{Acknowledgments}
KB gratefully acknowledges that this investigation was support by the National Institutes of Health under Ruth L. Kirschstein National Research Service Award \#5F32HG006411.

\bibliography{bibliography}
\bibliographystyle{mychicago}

%
\end{document}